\documentclass{paper}
\usepackage{amsfonts,amsmath,latexsym,amscd,graphicx,amssymb}

\begin{document}

\newcommand{\pa}{\partial}
\newcommand{\f}{\frac}
\newcommand{\st}{\stackrel}
\newcommand{\s}{\sqrt}



 \title{Variational derivation of the
Camassa-Holm shallow water equation}


\author{\normalsize Delia IONESCU-KRUSE\\
\normalsize Institute of Mathematics of
the Romanian Academy,\\
\normalsize P.O. Box 1-764, RO-014700, Bucharest,
 Romania\\
\normalsize E-mail: Delia.Ionescu@imar.ro\\[10pt]}

 \date{}

\maketitle
\begin{abstract}
\noindent We describe the physical hypothesis in which an
approximate model of water waves is obtained.  For an irrotational
 unidirectional shallow water flow, we derive the Camassa-Holm equation by a variational
 approach in the Lagrangian
formalism.
\end{abstract}
\section{Introduction}

 The Camassa-Holm equation reads

\begin{equation} U_t+2\kappa U_x+3UU_x-U_{txx}=2U_xU_{xx}+UU_{xxx}
\label{ch}\end{equation} with $x\in\mathbf{R}$, $t\in\mathbf{R}$,
$U(x,t)\in \mathbf{R}$. Subscripts here, and later, denote partial
derivatives. The constant $\kappa$ is related to the critical
shallow water speed.
 For
$\kappa$=0 the equation (\ref{ch}) possesses peaked soliton
solutions (\cite{holm}). The physical derivation of (\ref{ch}) as
a model for the evolution of a shallow water layer under the
influence of gravity, is due to Camassa and Holm \cite{holm}. See
also Refs. \cite{adrian1}, \cite{johnson1} for alternative
derivations within the shallow water regime. In \cite{misiolek} it
was shown that the equation (\ref{ch}) describes a geodesic flow
on the one dimensional central extension of the group of
diffeomorphisms  of the circle (for the case $\kappa$=0, a
geodesic flow on the diffeomorphism group of the circle, see also
\cite{adrian&boris}). It should be mentioned that, prior to
Camassa and Holm, Fokas and Fuchssteiner \cite{ff} obtained
formally, by the method of recursion operators in the context of
hereditary symmetries, families of integrable equations containing
(\ref{ch}). These equations are bi-Hamiltonian generalizations of
the KdV equation and possess infinitely many conserved quantities
in involution (\cite{ff}). In order to see how the equation
(\ref{ch}) relates with one of equations in the families
introduced in \cite{ff}, see \cite{f}, \S 2.2. Also, the equation
(5.3) in \cite{f2} is the equation (\ref{ch})  but with errors in
the coefficients.

The Camassa-Holm equation attracted a lot of interest, due to its
complete integrability \cite{CM} (for the periodic case),
\cite{cgi} and the citations therein (for the integrability on the
line), the existence of waves of permanent form and of breaking
waves \cite{CE} and the presence of peakon solutions \cite{holm}
(for a more rigorous study on the peakons see for example
\cite{bc}). An important question to be answered is  whether
solutions of the water wave problem can really be approximated by
solutions of the Camassa-Holm equation. In \cite{guido}  it is
shown that suitable solutions of the water wave problem and
solutions of the Camassa-Holm equation stay close together for
long times, in the case $\epsilon=\delta^2$, $\epsilon$ being the
amplitude parameter and $\delta$ the shallowness parameter.  In
\cite{guido} it is also shown that the peakon equation cannot
strictly be derived from the Euler equation and hence it is at
most a phenomenological model.

The present paper is concerned with the  physical hypothesis in
which  an approximate model of water waves is obtained, with the
derivation of the Camassa-Holm equation by a variational approach
in the Lagrangian formalism, and with the role of this equation
within the shallow water problem.

\section{The governing equations for water waves}

We consider water moving in a domain with a free upper surface at
$z=h_0+\eta(x,t)$, for a constant $h_0>0$, and a flat bottom at
$z=0$.  The undisturbed water surface is $z=h_0$. Let $(u(x,z,t),
v(x,z,t))$ be the velocity of the water - no motion take place in
the $y$-direction. The fluid is acted on only by the acceleration
of gravity $g$, the effects of surface tension are ignored. For
the gravity water waves, the appropriate equations of motion are
Euler's equations (EE )(see \cite{johnson-carte}). Another
realistic assumption for gravity water wave problem is the
incompressibility (constant density $\rho$) (see
\cite{johnson-carte}), which implies the equation of mass
conservation (MC). The boundary conditions for the water wave
problem are the kinematic boundary conditions as well as the
dynamic boundary condition. The kinematic boundary conditions
(KBC) express the fact that the same particles always form the
free water surface and that the fluid is assumed to be bounded
below by a hard horizontal bed $z=0$. The dynamic boundary
condition (DBC) express the fact that on the free surface the
pressure is equal with the constant atmospheric pressure denoted
$p_0$. Summing up, the exact solution for the water-wave problem
is given by the system \begin{equation}
\begin{array}{c}
\begin{array}{c}
u_t+uu_x+vu_z=-\f1{\rho} p_x\\  v_t+uv_x+vv_z=-\f1{\rho} p_z-g\\
\end{array}
\quad \quad \quad \quad \textrm{ (EE) }\\ \cr
 \qquad \qquad u_x+v_z=0  \qquad \qquad \qquad \qquad \textrm{ (MC)  }\\ \cr
\begin{array}{c}
  v=\eta_t+u\eta_x \, \, \textrm{ on }\,
z=h_0+\eta(x,t)\\
 v=0 \, \,
\textrm { on } z=0
\end{array}
\quad \textrm{ (KBC) }
\\ \cr
\qquad p=p_0, \,  \textrm{ on } z=h_0+\eta(x,t)
  \quad \qquad \textrm{ (DBC)} \end{array} \label{e+bc}
 \end{equation} where $p(x,z,t)$ denotes the pressure.

We non-dimensionalise this set of equations and boundary
conditions  using the undisturbed depth of water, $h_0$, as the
vertical scale,  a typical wavelength $\lambda$, as the horizontal
scale,  and a typical amplitude of the  surface wave $a$ (for more
details see \cite{johnson-carte}, \cite{johnson1}). An appropriate
choice for the scale of the horizontal component of the velocity
is $\sqrt{gh_0}$. Then, the corresponding time scale is
$\f\lambda{\sqrt{gh_0}}$ and the scale for the vertical component
of the velocity is $h_0\f{\sqrt{gh_0}}{\lambda}$. Thus, we define
the set of non-dimensional variables \begin{equation}
\begin{array}{c}
x\mapsto\lambda x,  \quad z\mapsto h_0 z, \quad \eta\mapsto a\eta,
\quad t\mapsto\f\lambda{\sqrt{gh_0}}t,\\
  u\mapsto  \sqrt{gh_0}u,
\quad v\mapsto h_0\f{\sqrt{gh_0}}{\lambda}v
\end{array} \label{nondim}\end{equation}
where, to avoid new notations, we have used the same symbols for
the non-dimensional variables  $x$, $z$, $\eta$, $t$, $u$, $v$, in
the right-hand side. The partial derivatives will be replaced by
\begin{equation}
\begin{array}{c}
u_t\mapsto \f{gh_0}{\lambda}u_t, \quad u_x\mapsto
\f{\sqrt{gh_0}}{\lambda}u_x, \quad u_z\mapsto\f {\sqrt{gh_0}}{h_0}u_z,\\

v_t\mapsto \f{gh_0^2}{\lambda^2}v_t, \quad v_x\mapsto
h_0\f{\sqrt{gh_0}}{\lambda^2}v_x, \quad v_z\mapsto\f {\sqrt{gh_0}}{\lambda}v_z,\\
\end{array}\label{derivate}\end{equation}

\noindent Let us now define the non-dimensional pressure. If the
water would be stationary, that is, $u\equiv v \equiv 0$, from the
first two equations and the last condition with $\eta=0$, of the
system (\ref{e+bc}), we get for a non-dimensionalised $z$, the
hydrostatic pressure $p_0+\rho g h_0(1-z)$. Thus, the
non-dimensional  pressure is defined  by \begin{equation} p\mapsto
p_0+\rho g h_0(1-z)+\rho g h_0 p \label{p}\end{equation} and
\begin{equation} p_x\mapsto \rho \f {gh_0}{\lambda} p_x, \quad
p_z\mapsto -\rho g+\rho g p_z\label{p'}\end{equation}

 Taking
into account (\ref{nondim}), (\ref{derivate}), (\ref{p}), and
(\ref{p'}), the water-wave problem (\ref{e+bc})  writes
 in
non-dimensional variables, as \begin{equation}
\begin{array}{c}
u_t+uu_x+vu_z=- p_x\\  \delta^2(v_t+uv_x+vv_z)=- p_z\\
 u_x+v_z=0\\
v=\epsilon(\eta_t+u\eta_x) \,   \textrm{ and } \,  p=\epsilon\eta
\, \, \textrm{ on }\,
z=1+\epsilon\eta(x,t)\\
 v=0 \, \,
\textrm { on } z=0
 \end{array}
\label{e+bc'} \end{equation}  where we have introduced the
amplitude parameter $\epsilon=\f a{h_0}$ and the shallowness
parameter $\delta=\f {h_0}{\lambda}$. The small-amplitude shallow
water is obtained in the limits $\epsilon\rightarrow 0$,
$\delta\rightarrow 0$. We observe that, on $z=1+\epsilon\eta$,
both $v$ and $p$ are proportional to $\epsilon$. This is
consistent with the fact that as $\epsilon\rightarrow 0$ we must
have $v\rightarrow 0$ and $p\rightarrow 0$, and it leads to the
following scaling of the non-dimensional variables
\begin{equation} p\mapsto \epsilon p,\quad
(u,v)\mapsto\epsilon(u,v) \label{scaling}\end{equation} where we
avoided again the introduction of a new notation. The problem
(\ref{e+bc'}) becomes \begin{equation}
\begin{array}{c}
u_t+\epsilon(uu_x+vu_z)=- p_x\\  \delta^2[v_t+\epsilon(uv_x+vv_z)]=- p_z\\
 u_x+v_z=0\\
  v=\eta_t+\epsilon u\eta_x  \, \textrm{ and } \,  p=\eta \, \, \textrm{ on }\,
z=1+\epsilon\eta(x,t)\\
 v=0 \, \,
\textrm { on } z=0
 \end{array}
\label{e+bc''} \end{equation}

Further, the parameter $\delta$ can be removed from the system
(\ref{e+bc''}) (see \cite{johnson1}), this being equivalent to use
only $h_0$ as the length scale of the problem. In order to do
this,  the non-dimensional variables $x$, $t$ and $v$ from
(\ref{nondim}) are replaced by
 \begin{equation} x\mapsto
\f{\sqrt{\epsilon}}{\delta}x,\quad t\mapsto
\f{\sqrt{\epsilon}}{\delta}t, \quad v\mapsto\f
{\delta}{\sqrt{\epsilon}}v\label{delta}\end{equation} Therefore
the equations in the system (\ref{e+bc''}) are recovered, but with
$\delta^2$ replaced by $\epsilon$ in the second equation of the
system, that is, this equation writes as  \begin{equation}
 \epsilon[v_t+\epsilon(uv_x+vv_z)]=- p_z\label{1}\end{equation}

\section{The Camassa-Holm equation}

The non-dimensionalisation and scaling presented above will be
useful in obtaining a scheme of approximation of the governing
water-wave problem (\ref{e+bc}).

In the limit $\epsilon\rightarrow 0$, that is for small-amplitude
waves, the system (\ref{e+bc''}) with the second equation given by
(\ref{1}) becomes \begin{equation}
\begin{array}{c}
u_t+p_x=0\\  p_z=0\\
 u_x+v_z=0\\
v=\eta_t \,  \textrm{ and } \,  p=\eta \, \, \textrm{ on }\,
z=1\\
 v=0 \, \,
\textrm { on } z=0
\end{array}
\label{small} \end{equation} From the second equation in
(\ref{small}), we get that $p$ does not depend on $z$. Because
$p=\eta(x,t)$ on $z=1$, we have \begin{equation} p=\eta(x,t) \,
\quad \textrm{ for any } \, \,  0\leq z\leq
1\label{2}\end{equation} Therefore, using the first equation in
(\ref{small}), we obtain
\begin{equation} u=-\int \eta_x(x,t)dt+\mathcal{F}(x,z)
\label{3}\end{equation} where $\mathcal{F}$ is an arbitrary
function. Differentiating (\ref{3}) with respect to $x$ and using
the third equation in (\ref{small}), we get, after an integration
against $z$, \begin{equation} v=z\int\eta_{xx}(x,t)dt
-\mathcal{G}(x,z)+\mathcal{G}(x,0)\label{4}\end{equation} where
$\mathcal{G}_z(x,z)= \mathcal{F}_x(x,z)$ and we have also taken
into account the last condition in the system (\ref{small}).
Making $z=1$ in (\ref{4}), and taking into account that $v=\eta_t$
on $z=1$, we get after a differentiation with respect to $t$, that
$\eta$ has to satisfy the equation \begin{equation}
\eta_{tt}-\eta_{xx}=0 \label{eta}\end{equation} The general
solution of this equation is $\eta(x,t)=f(x-t)+g(x+t)$, where $f$
and $g$ are differentiable functions.  It is convenient first to
restrict ourselves to waves which propagate in only one direction,
thus, we choose \begin{equation} \eta(x,t)=f(x-t)
\label{sol}\end{equation} Therefore, for $u$ and $v$ in (\ref{3}),
(\ref{4}) we get
\begin{equation} u=\eta+\mathcal{F}(x,z),\quad
v=-z\eta_x-\mathcal{G}(x,z)+\mathcal{G}(x,0)\end{equation} with
$\mathcal{G}_z(x,z)= \mathcal{F}_x(x,z)$,
$\mathcal{G}(x,1)=\mathcal{G}(x,0)$, arbitrary functions. Thus,
the solutions to the shallow water problem are determined by the
evolution of the function $\eta(t,x)$, which represents the
displacement of the free surface from the undisturbed (flat)
state.

\vspace{0.5cm}

\textit{Under the assumption that the fluid is irrotational, we
get} \begin{equation} \mathcal{F}(x,z)=\textrm{ const}:=c_0 \quad
\textrm{ and } \quad \mathcal{G}(x,z)=0\end{equation}

\vspace{0.5cm}

\noindent Indeed, if the fluid is irrotational the vorticity is
zero, that is, in addition to the system (\ref{e+bc}), we also
have the equation \begin{equation} u_z-v_x=0\label{vorticity}
\end{equation}
For a discussion of the role of vorticity in water wave flows see
for example \cite{CS}, \cite{johnson2}.

\noindent In the equation (\ref{vorticity}), the velocity
components $u$ and $v$ are written in the physical (dimensional)
variables. If we non-dimensionalise this equation using
(\ref{nondim}), (\ref{derivate}), we obtain
\begin{equation} u_z=\delta^2 v_x \label{omega}\end{equation}
After scaling (\ref{scaling}) and transformation (\ref{delta}),
the equation (\ref{omega}) writes as
\begin{equation} u_z=\epsilon v_x \end{equation} Therefore, in the limit
$\epsilon\rightarrow 0$, we get in addition to the system
(\ref{small}), the equation \begin{equation} u_z=0
\label{5}\end{equation} The relation (\ref{2}) remains the same
but instead of (\ref{3}) we have now \begin{equation} u=-\int
\eta_x(x,t)dt+\tilde{\mathcal{F}}(x) \label{3'}\end{equation}
where $\tilde{\mathcal{F}}$ is an arbitrary function. Using the
third equation in (\ref{small}), we get now \begin{equation}
v=-zu_x=z\left(\int\eta_{xx}(x,t)dt
-\tilde{\mathcal{F}}'(x)\right)\label{4'}\end{equation} where we
have  taken into account the last condition in the system
(\ref{small}). Making $z=1$ in (\ref{4'}), and taking into account
that $v=\eta_t$ on $z=1$, we get after a differentiation with
respect to $t$, that $\eta$ has to satisfy the equation
(\ref{eta}). We consider the solution of (\ref{eta}) into the form
the solution (\ref{sol}). Therefore, for $u$ and $v$ in
(\ref{3'}), (\ref{4'}), we have $ u=\eta+\tilde{\mathcal{F}}(x),$
$v=-z\left(\eta_x+\tilde{\mathcal{F}}'(x)\right)$. The condition
$v=\eta_t$ on $z=1$, yields
$\tilde{\mathcal{F}}(x)=\textrm{const}:=c_0$. Thus, for the
irrotational case the solution of the system (\ref{small}) plus
the equation (\ref{5}), can be written into the form
\begin{equation} \eta(x,t)=f(x-t), \quad u=\eta+c_0,\quad
v=-z\eta_x\label{solirrot}\end{equation} We underline the fact
that in our approximation the vertical velocity component
maintains a dependence on the $z$-variable.  In analyzing the
motion of the fluid particles, this means that the particles below
the surface may perform a vertical motion. This is in agreement
with a recent general result obtained in \cite{c2007} for the
Stokes waves, i.e. particular waves in an irrotational flow which
are solutions of the full Euler equations.

By consistently neglecting the $\epsilon$ contribution, we will
derive using variational methods in the Lagrangian formalism (see
\cite{adrian1}), the equation (\ref{ch}) governing unidirectional
propagation of shallow water waves.

In the Lagrangian formulation of a fluid, the flow pattern is
obtained by describing the path of each individual water particle.
Consider the ambient space $M$ whose points are supposed to
represent the fluid particles at $t=0$. A diffeomorphism of $M$
represents the rearrangement of the particles with respect to
their initial positions. The motion of the fluid is described by a
time-dependent family of orientation-preserving diffeomorphisms
$\gamma(t,\cdot)\in$ Diff($M$).  A point $x$ in $M$
 follows the trajectory $\gamma(t,x)$ through $M$.
For our problem, since a particle on the water's free surface will
always stay on the surface and describes progressive plane wave
(no motion take place in the $y$ direction), we may regard the
motion at that of a one-dimensional
 membrane.
 For the
 one-dimensional periodic
 motion, $M=\mathbf{S}^1$ the unit circle. We can allow $M=\mathbf{R}$
 and add the technical assumption that the
 smooth functions defined on  $\mathbf{R}$ with value in $\mathbf{R}$
  vanish rapidly at $\pm\infty$ together
 with as many derivatives as necessary (see \cite{c} for a possible choice of weighted Sobolev
spaces). In what follows we focus on the latter
 situation.

 For a fluid particle initially located at $x$, the velocity at
 time $t$ is  $\gamma_t(t,x)$, this being the material velocity
 used in the Lagrangian description. The spatial velocity, used
 in the Eulerian description, is the flow velocity
 $w(t,X)=\gamma_t(t,x)$ at the location  $X=\gamma(t,x)$ at time
 $t$,
 that is, $w(t,\cdot)=\gamma_t\circ\gamma^{-1}$. In Lagrangian description,
 the equation of motion is the equation satisfied by a critical
 point of a certain functional (called the action) defined
 on all paths  $\{\gamma(t,\cdot),$ $t\in[0,T]\}$ in $\textrm{Diff}(\mathbf{R})$,
  having fixed endpoints.
   Following Arnold's approach to Euler equations on diffeomorphism
groups (\cite{arnold}),  the action for our problem
   will be obtained by transporting
  the kinetic energy to all  tangent spaces of
  Diff$(\mathbf{R})$ by means of right translations.
  For small surface elevation, the potential energy is
   negligible compared to the kinetic energy.
   Taking into account (\ref{solirrot}), the kinetic energy on the surface is
\begin{equation} K=\f 1{2}\int_{-\infty}^{\infty}(u^2+v^2)dx=\f
1{2}\int_{-\infty}^{\infty} \left[u^2+(1+\epsilon
\eta)^2u_x^2\right]dx\approx \f 1{2}\int_{-\infty}^{\infty}
\left(u^2+u_x^2\right)dx
   \end{equation}
to the order of our approximation (see \cite{adrian1}). We observe
that if we replace the path $\gamma(t,\cdot)$ by
$\gamma(t,\cdot)\circ\psi(\cdot)$, for a fixed time-independent
$\psi$ in Diff($\mathbf{R}$), then the spatial velocity is
unchanged $\gamma_t\circ \gamma^{-1}$. Transforming $K$ to a
right-invariant Lagrangian, the action on a path
$\gamma(t,\cdot)$, $t\in [0,T]$, in Diff($\mathbf{R}$) is
\begin{equation} \mathfrak{a}(\gamma)= \f
1{2}\int_0^T\int_{-\infty}^{\infty}
\{(\gamma_t\circ\gamma^{-1})^2+[\partial_x(\gamma_t\circ\gamma^{-1})]^2\}dxdt\label{action}\end{equation}
The critical points of the action (\ref{action}) in the space of
paths with fixed endpoints, verify \begin{equation}
\f{d}{d\varepsilon} \mathfrak{a}(\gamma+\varepsilon\varphi)\Big
|_{\varepsilon=0}=0,\label{critic}\end{equation} for every path
$\varphi(t,\cdot)$, $t\in[0,T]$, in $\textrm{Diff}(\mathbf{R})$
with endpoints at zero, that is,
$\varphi(0,\cdot)=0=\varphi(T,\cdot)$ and such that
$\gamma+\varepsilon\varphi$ is a small variation of $\gamma$ on
Diff($\mathbf{R}$). Taking into account (\ref{action}), the
condition (\ref{critic}) becomes \begin{eqnarray}
\int_0^T\int_{-\infty}^{\infty}&&\left\{\left(\gamma_t\circ\gamma^{-1}\right)
\f{d}{d\varepsilon}\Big
|_{\varepsilon=0}\left[(\gamma_t+\varepsilon\varphi_t)\circ(\gamma+\varepsilon\varphi)^{-1}\right
] \right.\nonumber\\
&&\left. +
\partial_x(\gamma_t\circ\gamma^{-1})\f{d}{d\varepsilon}\Big
|_{\varepsilon=0}\left[\partial_x\left((\gamma_t+\varepsilon\varphi_t)
\circ(\gamma+\varepsilon\varphi)^{-1}\right)\right]\right\}dxdt=0\nonumber\\
\label{6}\end{eqnarray} After calculation (see \cite{adrian1}), we
get \begin{eqnarray} \hspace{-0.6cm}\f{d}{d\varepsilon}\Big
|_{\varepsilon=0}\left[(\gamma_t+\varepsilon\varphi_t)\circ(\gamma+\varepsilon\varphi)^{-1}\right
]=&&\hspace{-0.5cm}\varphi_t\circ\gamma^{-1}-(\varphi\circ\gamma^{-1})\partial_x(\gamma_t\circ\gamma^{-1})
\nonumber\\
=&&\hspace{-0.5cm}\partial_t(\varphi\circ\gamma^{-1})+(\gamma_t\circ\gamma^{-1})\partial_x(\varphi\circ\gamma^{-1})\nonumber\\
&&\hspace{-0.5cm} -
(\varphi\circ\gamma^{-1})\partial_x(\gamma_t\circ\gamma^{-1})
\label{11}\end{eqnarray} and
 \begin{eqnarray}  \hspace{-0.6cm}\f{d}{d\varepsilon}\Big
|_{\varepsilon=0}\left[\partial_x\left((\gamma_t+\varepsilon\varphi)\circ(\gamma+\varepsilon\varphi)^{-1}\right)\right]
&=&
\partial_x(\varphi_t\circ\gamma^{-1})-\partial_x(\gamma_t\circ\gamma^{-1})\partial_x(\varphi\circ\gamma^{-1})\nonumber\\
&&-(\varphi\circ\gamma^{-1})\partial_x^2(\gamma_t\circ\gamma^{-1})
\nonumber\\
&=&\partial_{tx}(\varphi\circ\gamma^{-1})+(\gamma_t\circ\gamma^{-1})\partial^2_x(\varphi\circ\gamma^{-1})\nonumber\\
&&-(\varphi\circ\gamma^{-1})\partial_x^2(\gamma_t\circ\gamma^{-1})\label{12}\end{eqnarray}
where are used the formulas of the type \begin{equation} \f
{d}{d\varepsilon}\Big
|_{\varepsilon=0}\left[(\gamma+\varepsilon\varphi)^{-1}\right]=-\f{\varphi\circ\gamma^{-1}}{\gamma_x\circ\gamma^{-1}}\label{7}\end{equation}
\begin{equation}
\partial_x(\gamma_t\circ\gamma^{-1})=\f{\gamma_{tx}\circ\gamma^{-1}}{\gamma_x\circ\gamma^{-1}}
\end{equation} \begin{equation}
\partial_t(\varphi\circ\gamma^{-1})=\varphi_t\circ\gamma^{-1}+\left(\varphi_x\circ\gamma^{-1}\right)\partial_t(\gamma^{-1})=
\varphi_t\circ\gamma^{-1}-(\gamma_t\circ\gamma)\partial_x(\varphi\circ\gamma^{-1})
\end{equation}
Thus, denoting $\gamma_t\circ\gamma^{-1}=u$, from (\ref{11}),
(\ref{12}), the condition (\ref{6}) writes as \begin{eqnarray}
\int_0^T\int_{-\infty}^{\infty}&&
\left\{u\left[\partial_t(\varphi\circ\gamma^{-1})+u\partial_x(\varphi\circ\gamma^{-1})
-(\varphi\circ\gamma^{-1})u_x\right]\right.\nonumber\\
&& \left.+ u_x\left[
\partial_{tx}(\varphi\circ\gamma^{-1})+u\partial^2_x(\varphi\circ\gamma^{-1})
-(\varphi\circ\gamma^{-1})u_{xx}\right]\right\}dxdt
=0\nonumber\\
\label{10}\end{eqnarray} We integrate by parts with
respect to $t$ and $x$ in the above formula, we take into account
that $\varphi$ has endpoints at zero, the smooth functions defined
on $\mathbf{R}$ with value in $\mathbf{R}$, together
 with as many derivatives as necessary, vanish rapidly at $\pm\infty$, and we obtain \begin{equation}
-\int_0^T\int_{-\infty}^\infty(\varphi\circ\gamma^{-1})\left[u_t+3uu_x
-u_{txx}- 2u_xu_{xx}-uu_{xxx}\right]dxdt=0 \end{equation}

\vspace{0.5cm}

\noindent \textit{Therefore, we get that for an irrotational
 unidirectional shallow water flow,   the horizontal velocity component of the water $u(x,t)$ satisfies the Cammassa-Holm
equation (\ref{ch}) for $\kappa=0$}.

\vspace{0.5cm}

Let us see now which equation fulfill the displacement $\eta(x,t)$
of the free surface from the flat state. Taking into account
(\ref{solirrot}), the kinetic energy on the surface is
\begin{eqnarray} K=\f 1{2}\int_{-\infty}^{\infty}(u^2+v^2)dx&=&\f
1{2}\int_{-\infty}^{\infty} \left[(\eta+c_0)^2+(1+\epsilon
\eta)^2\eta_x^2\right]dx\nonumber\\
&\approx& \f 1{2}\int_{-\infty}^{\infty}
\left[\left(\eta+c_0\right)^2+\eta_x^2\right]dx
   \end{eqnarray}
to the order of our approximation. Transforming $K$ to a
right-invariant Lagrangian, the action on a path
$\Gamma(t,\cdot)$, $t\in [0,T]$, in Diff($\mathbf{R}$) is
\begin{equation} \mathfrak{a}(\Gamma)= \f
1{2}\int_0^T\int_{-\infty}^{\infty}
\{(\Gamma_t\circ\Gamma^{-1}+c_0)^2+[\partial_x(\Gamma_t\circ\Gamma^{-1})]^2\}dxdt\label{action1}\end{equation}
The critical points of the action (\ref{action1}) in the space of
paths with fixed endpoints, verify \begin{equation}
\f{d}{d\varepsilon} \mathfrak{a}(\Gamma+\varepsilon\Phi)\Big
|_{\varepsilon=0}=0,\label{critic1} \end{equation} for every path
$\Phi(t,\cdot)$, $t\in[0,T]$, in $\textrm{Diff}(\mathbf{R})$ with
endpoints at zero, that is, $\Phi(0,\cdot)=0=\Phi(T,\cdot)$ and
such that $\Gamma+\varepsilon\Phi$ is a small variation of
$\Gamma$ on Diff($\mathbf{R}$). Taking into account
(\ref{action1}), the condition (\ref{critic1}) becomes
\begin{eqnarray}
\int_0^T\int_{-\infty}^{\infty}&&\left\{\left(\Gamma_t\circ\Gamma^{-1}+c_0\right)
\f{d}{d\varepsilon}\Big
|_{\varepsilon=0}\left[(\Gamma_t+\varepsilon\Phi_t)\circ(\Gamma+\varepsilon\Phi)^{-1}\right
] \right.\nonumber\\
&&\left. +
\partial_x(\Gamma_t\circ\Gamma^{-1})\f{d}{d\varepsilon}\Big
|_{\varepsilon=0}\left[\partial_x\left((\Gamma_t+\varepsilon\Phi_t)
\circ(\Gamma+\varepsilon\Phi)^{-1}\right)\right]\right\}dxdt=0\nonumber\\
\label{66}\end{eqnarray} After calculation, denoting
$\Gamma_t\circ\Gamma^{-1}=\eta$, (\ref{66}) writes as
\begin{eqnarray} \int_0^T\int_{-\infty}^{\infty}&& \left\{(\eta
+c_0)\left[\partial_t(\Phi\circ\Gamma^{-1})+\eta\partial_x(\Phi\circ\Gamma^{-1})
-(\Phi\circ\Gamma^{-1})\eta_x\right]\right.\nonumber\\
&& \left.+ \eta_x\left[
\partial_{tx}(\Phi\circ\Gamma^{-1})+\eta\partial^2_x(\Phi\circ\Gamma^{-1})
-(\Phi\circ\Gamma^{-1})\eta_{xx}\right]\right\}dxdt=0\nonumber\\
\label{10'}\end{eqnarray} We integrate by parts with respect to
$t$ and $x$ in the above formula, we take into account that $\Phi$
has endpoints at zero, the smooth functions defined on
$\mathbf{R}$ with values in $\mathbf{R}$, together
 with as many derivatives as necessary, vanish rapidly at $\pm\infty$, and we obtain \begin{equation}
-\int_0^T\int_{-\infty}^\infty(\Phi\circ\Gamma^{-1})\left[\eta_t+3\eta\eta_x
+2c_0\eta_x-\eta_{txx}-
2\eta_x\eta_{xx}-\eta\eta_{xxx}\right]dxdt=0 \end{equation}

\vspace{0.5cm}

 \noindent \textit{Therefore, we get that for an
irrotational
 unidirectional shallow water flow,   the displacement $\eta(x,t)$ of the free
surface from the flat state, satisfies the Camassa-Holm equation
(\ref{ch}) for $\kappa=c_0$.}

\noindent \textbf{Acknowledgments} I would like to thank Prof. A.
Constantin  for many interesting and useful discussions on the
subject of water waves and for his kind hospitality at Trinity
College  Dublin.

\label{lastpage}
\end{document}